\documentclass[twocolumn,showpacs,preprintnumbers,amsmath,amssymb]{revtex4}

\usepackage{graphicx}
\usepackage{dcolumn}
\usepackage{bm}

\newfont{\Fr}{eufm10 scaled\magstep1}
\newfont{\Fra}{eufm10 scaled\magstephalf}

\def\A&A#1#2{{ Astron. Astrophys.} {\bf #1}, #2}

\def\PR#1#2{{ Phys. Rev.} {\bf #1}, #2}
\def\PL#1#2{{ Phys. Lett.} {\bf #1}, #2}

\def\NPA#1#2{{ Nucl. Phys.} {\bf #1}, #2}

\begin{document}

\preprint{APS/123-QED}

\title{Photo-disintegration cross section measurements on 
$^{186}$W, $^{187}$Re and $^{188}$Os: implications to the Re-Os 
cosmochronology}

\author{
T. Shizuma$^1$, 
H. Utsunomiya$^2$, 
P. Mohr$^3$,
T. Hayakawa$^1$, 
S. Goko$^2$, A. Makinaga$^2$, H. Akimune$^2$, T. Yamagata$^2$, M. Ohta$^2$,
H. Ohgaki$^4$,
Y.-W. Lui$^5$,
H. Toyokawa$^6$, A. Uritani$^7$, and 
S. Goriely$^8$
}

\affiliation{
$^1$Advanced Photon Research Center, Japan Atomic Energy Research Institute, Japan\\
$^2$Department of Physics, Konan University, Japan\\
$^3$ Strahlentherapie, Diakoniekrankenhaus Schw\"abisch Hall, Germany \\
$^4$Institute of Advanced Energy, Kyoto University, Japan\\
$^5$Cyclotron Institute, Texas A\&M University, USA\\
$^6$Photonics Research Institute, National Institute of Advanced Industrial 
Science and Technology, Japan\\
$^7$National Metrology Institute, National Institute of Advanced Industrial 
Science and Technology, Japan\\
$^8$Institute d'Astronomie et d'Astrophysique, Universit\'{e} Libre de 
Bruxelles, Belgium\\
}

\date{\today}

\begin{abstract}
Cross sections of the $^{186}$W, $^{187}$Re, $^{188}$Os($\gamma,n$) 
reactions were measured using quasi-monochromatic photon beams
from laser Compton scattering (LCS) with average energies 
from 7.3 to 10.9 MeV. The results are compared with the predictions 
of Hauser-Feshbach statistical calculations using four different 
sets of input parameters. In addition, the inverse neutron capture cross sections
were evaluated by constraining the model parameters, especially 
the $E1$ strength function, on the basis of the experimental 
data. The present experiment helps to further constrain the correction factor 
$F_{\sigma}$ for the neutron capture on the 9.75 keV state in $^{187}$Os. 
Implications of $F_{\sigma}$ to the Re-Os cosmochronology 
are discussed with a focus on the uncertainty in the estimate of 
the age of the Galaxy.   
\end{abstract}

\maketitle


\section{Introduction}

Owing to the long half-life of $^{187}$Re, the $^{187}$Re-$^{187}$Os 
pair may serve as a cosmochronometer to measure the duration of stellar 
nucleosynthesis that precedes the solidification of the solar system \cite{Clay64}.
By adding the age of the solar system ($\sim$4.6 Gyr), 
it provides the age of the Galaxy. The facts that both $^{186}$Os 
and $^{187}$Os are produced only 
by the s-process nucleosynthesis apart from the cosmoradiogenic 
yield of $^{187}$Os and that the isotopic solar abundance ratio 
of $^{186}$Os and $^{187}$Os \cite{AnGr89} is available make this 
chronometer potentially reliable in the sense that it is independent of r-process models. 
The quantitative interpretation is, however, complicated by the possible enhancement of
$^{187}$Re-$^{187}$Os transmutation rates in stellar condition,
the stellar production and destruction of $^{187}$Re and $^{187}$Os
during the chemical evolution of the Galaxy,
the possible existence of s-process branchings at $^{185}$W
and $^{186}$Re, and the neutron capture
by the 9.75 keV first excited state in $^{187}$Os
\cite{Yoko83,Arno84,Taka03,Woos79}.

The last issue on the effect of neutron capture on the 9.75 keV 
state in $^{187}$Os was raised in Ref. \cite{Woos79}. In the local approximation, 
the ratio of the s-process yields of $^{186}$Os and $^{187}$Os, 
$N_s(^{186}{\rm Os})/N_s(^{187}{\rm Os})$, can be expressed as 

\begin{equation}
\frac{N_s(^{187}{\rm Os})}{N_s(^{186}{\rm Os})} \approx 
F_\sigma \frac{\langle\sigma\rangle({\rm ^{186}Os})}
{\langle\sigma\rangle({\rm ^{187}Os})}.
\label{localapprox}
\end{equation}

\noindent
Here, $\langle\sigma\rangle({\rm ^{186}Os})$ and 
$\langle\sigma\rangle({\rm ^{187}Os})$ are the Maxwellian-averaged 
neutron capture cross sections on $^{186}$Os and $^{187}$Os
in the ground states, respectively. The $F_\sigma$ value accounts for the 
correction to the cross section due to the neutron capture on 
the 9.75 keV state in $^{187}$Os which is substantially populated 
at typical s-process temperatures $T\simeq$1-3$\times10^8$ K. 
It is defined by  

\begin{equation}
F_\sigma=\frac{\langle\sigma\rangle({\rm ^{187}Os})}
{\langle\sigma\rangle^*({\rm ^{187}Os})}
\label{eq_fsig}
\end{equation}

\noindent
with ${\langle\sigma\rangle^*(^{187}{\rm Os})}$ being the 
Maxwellian-averaged neutron capture cross section on $^{187}$Os 
at a given stellar temperature.  Here, the first excited state 
in $^{186}$Os and the second excited state in $^{187}$Os 
lie at higher excitation energies of 137 keV and 74 keV, respectively, 
and therefore their contributions to neutron capture 
in the stellar condition may safely be ignored.  
In 1970s, it was of critical concern whether or not the $F_\sigma$ 
value exceeds unity, because it has a great impact on the age of the Galaxy; 
the larger the  $F_\sigma$, the smaller the age. However, there was a 
large spread in the early estimate (0.80-1.10 \cite{Woos79}, 0.8 \cite{Holm76}, 
and $\sim$1.5 \cite{Fowl73}). 

This concern, combined with the fact that a direct measurement of neutron 
capture on the 9.75 keV state is virtually impossible, led to measurements 
of neutron capture on $^{186,187,188}$Os \cite{Wint80,Wint82} and neutron 
inelastic scattering to the 9.75 keV state in $^{187}$Os \cite{Hers83,Mack83}. 
The measured neutron capture cross sections were in good agreement with 
those of the earlier measurements \cite{BrLS76,Brow81}. But, the statistical analysis 
of the capture data gave a lower bound of 0.30 $b$ to the inelastic scattering 
cross section $\sigma_{nn'}$ at a neutron energy of 30 keV, and 
by combining with an upper limit of $\sigma_{nn'}=0.5$ $b$ \cite{Wint74}, 
gave $F_\sigma\sim1$. In contrast, the two neutron inelastic 
scattering data, $\sigma_{nn'}=1.13\pm0.2$ $b$ at 60 keV \cite{Hers83} 
and $1.5\pm0.2$ $b$ at 34 keV \cite{Mack83}, are consistent with 
$F_\sigma=$0.80-0.83 and 0.80, respectively, within the statistical models.  

More recent efforts have been made toward a unified 
statistical model analysis of all available data including 
measurements of ($n,\gamma$) cross sections \cite{WiMH87} and 
elastic/inelastic scattering cross sections \cite{McEl89} on a 
neighboring nucleus $^{189}$Os in which the ground state 
with $J^{\pi}$ = 3/2$^-$ and the first excited state with 
1/2$^-$ at 36 keV appear in the reverse order of the corresponding 
states in $^{187}$Os. These analyses showed that $F_\sigma$=0.79-0.83.   

In the Hauser-Feshbach statistical model calculations, however, 
large uncertainties may arise from the $\gamma$-ray transmission 
coefficients rather than the neutron optical potential and the 
level density. Information on the $\gamma$-ray transmission 
coefficients for neutron captured states in the low-energy tail 
of the giant dipole resonance in $^{188}$Os can be obtained 
in the inverse photo-disintegration of $^{188}$Os. But, the 
conventional Lorentzian model based on the previous photo-disintegration 
data on $^{188}$Os \cite{Ber69} may not be satisfactory for two 
reasons. One reason is that the data taken with $\gamma$-ray beams 
from the positron annihilation in flight exhibit non-vanishing cross 
sections even below the neutron threshold (see sec. \ref{sec-cross}). 
The non-vanishing cross sections 
may be attributed to contributions from the positron bremsstrahlung. 
The other is that microscopic models can predict the $E1$ $\gamma$ 
strength function more reliably than the Lorentzian model.  

Besides $F_\sigma$, the effect of the s-process branchings at 
$^{185}$W and/or $^{186}$Re was parameterized as $F_b$ 
and investigated within the framework of the schematic s-process 
models \cite{Arno84}. More recently, neutron capture cross sections 
were measured for neighboring $^{185}$Re and $^{187}$Re 
nuclei to derive the statistical model parameters from a 
consistent systematics \cite{Kaep91}. With the improved 
parameters for s-process analysis, a stellar model calculation for 
low-mass AGB stars showed that the local approximation was disturbed 
by the branchings at $^{185}$W and $^{186}$Re. However, 
the precise physical conditions of the AGB model need to be 
scrutinized before any definite conclusion on $F_b$ is drawn. 
This is clearly beyond the scope of the present study. Instead, the relevant 
statistical parameters, particularly the $E1$ $\gamma$ strength 
function, can further be improved by the photo-disintegration 
measurement on $^{186}$W.  

In the present study, we have measured photoneutron cross sections
of $^{186}$W, $^{187}$Re and $^{188}$Os using tunable 
quasi-monochromatic $\gamma$-ray beams from laser Compton 
scattering (LCS). The photo-disintegration data allow us to constrain 
$F_\sigma$ values within the Hauser-Feshbach 
statistical model and discuss their implications to the Re-Os cosmochronology.  
Part of the present data has already been published \cite{Mohr04}.

\section{Experimental procedure}
Photoneutron cross section measurements on 
$^{186}$W, $^{187}$Re and $^{188}$Os were performed at the 
National Institute of Advanced Industrial Science and 
Technology (AIST). Tunable quasi-monochromatic photon beams 
were generated by Compton scattering of laser photons, 
with relativistic electrons circulating in the 
storage ring TERAS \cite{Ohg91}. A Nd:YLF Q-switch laser at a wavelength 
of 527 nm in second harmonics was operated at a frequency of 2 kHz.
The electron energy was varied in the range from 450 to 588 MeV
to produce LCS photons with the average energy from 7.3 and 10.9 MeV.
A 20 cm lead collimator with a small hole of 2 mm in diameter 
was placed at approximately 6 m downstream from the interaction area 
which defines a scattering cone of the LCS photons. 
The typical energy resolution was 10 \% in FWHM. Further details on 
the experimental setup can be found in Ref. \cite{Utsu03}.

\begin{figure}
\includegraphics[scale=0.5]{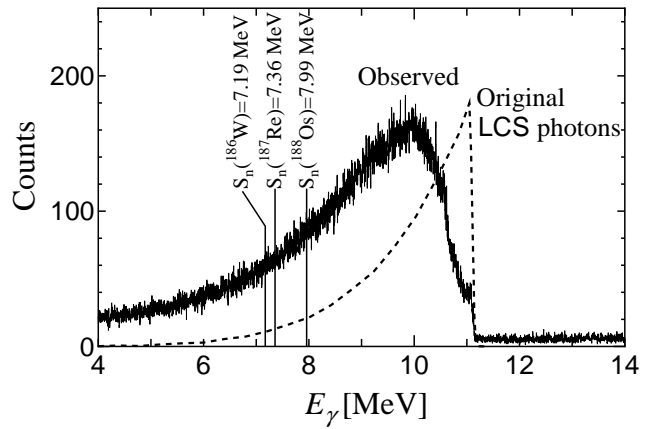}
\caption{An observed spectrum of the LCS photons measured 
with an HP-Ge detector (solid line). 
The original LCS photons simulated by the Monte Carlo 
code EGS4 is also shown with dashed line. The neutron separation
energies $S_n$ of $^{186}$W, $^{187}$Re and $^{188}$Os are 
indicated by solid lines. Fractional photons with 
energies higher than $S_n$ are responsible for the 
photoneutron reactions.}
\label{lcs}
\end{figure}

Figure \ref{lcs} shows an energy spectrum of the LCS photons 
measured by an HP-Ge detector with a relative efficiency of 120 \%. 
The energy was calibrated at 1460.8 and 2614.5 keV 
with $^{40}$K and $^{208}$Tl radioactive isotopes of natural origin. 
A Monte Carlo simulation was performed with the EGS4 code \cite{Nel85} 
to analyze the response of the Ge detector. The energy distribution of 
incident LCS photons (dotted line in Fig. \ref{lcs}) was determined 
so as to reproduce the observed response function (solid line in 
Fig. \ref{lcs}). Photons with energies higher than the neutron 
threshold ($S_n$=7.19 MeV for $^{186}$W, $S_n$=7.36 MeV for $^{187}$Re 
and $S_n$=7.99 MeV for $^{188}$Os) are responsible for the ($\gamma$,$n$) 
reactions. The fraction of these photons in the total photon flux 
and the average photon energy were obtained from the original LCS 
photon spectrum. 

The beam current of the electron storage ring decreases exponentially in a normal condition with a lifetime $\sim$ 6 hours.  In the EGS4 Monte Carlo simulation, it was found that the electron beam size in the region of the interaction with laser photons varied with time: for example, from 2.2 mm in diameter at 166 mA to 1.2 mm at 72 mA in the $^{187}$Re measurement.  A space-charge effect is considered to be a main cause for the decrease in the beam size.  This beam size effect, which was evident in long runs near the neutron thresholds, introduced uncertainties in the fraction of the LCS photon beam above the threshold.  The resultant uncertainty was estimated to be 1 - 6 \% in the present experiment.  On the other hand, the average photon energy was determined well within 40 keV.

\begin{figure}
\includegraphics[scale=0.5]{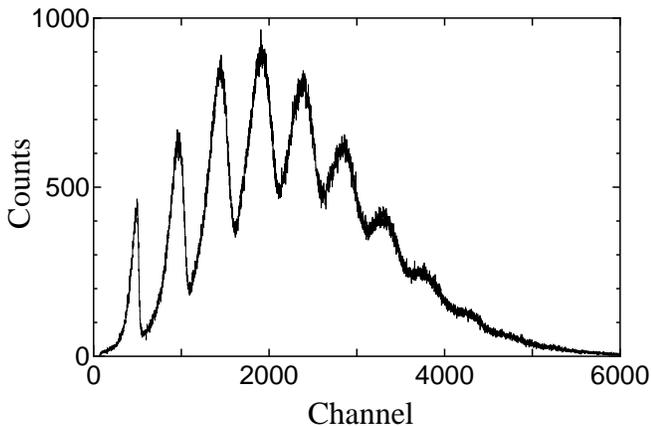}
\caption{Pile-up spectrum of the LCS photons measured with a large
volume (8"$\times$12") NaI(Tl) scintillation detector. In this spectrum 
the average photon number per laser pulse is 5.4 which leads to 
the LCS photon intensity of 1.08$\times10^4$ /s.}
\label{nai}
\end{figure}

The number of LCS photons was monitored during the experiment
using a large volume (8"$\times$12") NaI(Tl) scintillation detector 
placed behind targets. A typical pile-up spectrum is shown in 
Fig. \ref{nai}. The pulse height of the spectrum is proportional to 
the number 
of LCS photons per beam pulse. The photon flux was determined with 
3 \% uncertainty based on a statistical analysis on the 
pile-up spectrum \cite{Toyo00}.

Metallic powders of 1246 mg $^{186}$W, 996 mg $^{187}$Re, and 693 mg 
$^{188}$Os enriched to 99.79, 99.52 \% and 94.99 \%, respectively, 
were pressed to self-supporting tablets with a diameter of 8 mm. 
The $^{188}$Os powder included major contaminants of $^{189}$Os 
(2.55 \%), $^{190}$Os (1.27 \%) and $^{192}$Os (0.97 \%). 
These tablets were mounted inside thin containers made of aluminum, and
were irradiated with the LCS photon beams. The threshold energy of the 
$(\gamma,n)$ reaction on $^{27}$Al is 13.06 MeV, which is 
higher than those for $^{186}$W, $^{187}$Re and $^{188}$Os. 
The present photoneutron cross section measurements were performed 
at energies below the threshold energy and therefore undisturbed by the 
$^{27}$Al$(\gamma,n)$ reaction. Further, measurements with an empty aluminum 
container (blank target) showed that no background neutrons were produced from photo-disintegration of possible impurities in the aluminum.

\begin{figure}
\includegraphics[scale=0.5]{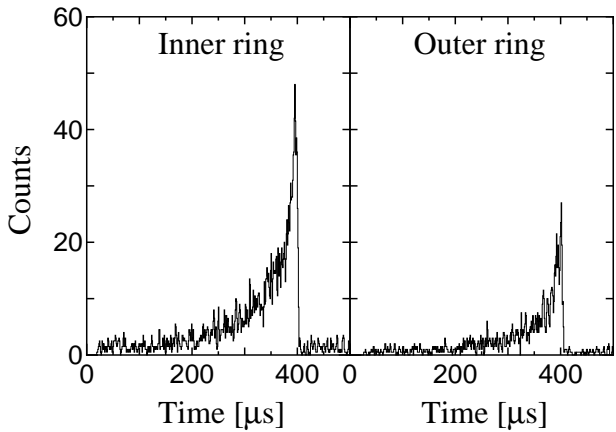}
\caption{Typical time-correlated spectra between laser pulses 
and neutrons detected in the inner (left panel) and outer (right panel) 
rings.  The time correlation was measured by using a TAC module with the neutron
signal and the 2 kHz laser pulse being the start and stop, respectively. }
\label{tac}
\end{figure}

Emitted neutrons were detected by sixteen $^3$He proportional 
counter (EURISYS MESURES 96NH45) embedded into a polyethylene 
moderator. Two sets of eight counters were placed in 
double concentric (inner and outer) rings at 7 and 10 cm 
from the beam axis. Time correlations (Fig. \ref{tac}) between 
the neutron signal and the laser pulse were measured to estimate the 
number of background neutrons that arrived randomly at the $^3$He 
detectors. These background neutrons were most likely produced by 
bremsstrahlung arising from collisions of electrons
with residual gaseous molecules in the storage ring.
In the moderation time distribution, constant events above 400 $\mu$s and at small correlation times were taken to be background neutrons.  The constant background was further confirmed by using a 1 kHz laser and a wider (1 ms) time range (see, for example, 
Fig. 3 of Ref. \cite{Hara03}).  The background subtraction is included in the statistical uncertainties through the error propagation.  

The neutron detection efficiency was measured at the average neutron 
energy 2.14 MeV with a standard $^{252}$Cf source. The dependence of 
the efficiency on neutron energy was determined by a Monte Carlo 
MCNP simulation with the statistical accuracy less than 0.5 \% 
\cite{Hara03}. The so-called ring ratio between the 
count rates of inner and outer rings was used to determine 
the average energy of emitted neutrons \cite{Utsu03}. 
A polynomial fit to the energy dependence of the ring ratios was made as 
in \cite{Utsu03}, where an asymptotic value of the ring ratio (8.0) 
at 10 keV simulated for the present neutron detector \cite{Hara03} 
was used as a constraint. 
The ring ratio varied between 7.4 and 2.9, indicating the 
neutron energies of several tens to a few hundreds keV 
for measurements close above the thresholds and up to 
0.82 MeV at higher photon energies.
The uncertainty in the neutron energy thus 
determined was estimated to be 10-15 keV, resulting in the uncertainty 
less than 0.8 \% in the total neutron detection efficiency. Note that 
the total efficiency is nearly constant (44.4-44.3 \%) over the neutron 
energy from 1 MeV to 400 keV and that it slowly decreases to 39.5 \% 
at 50 keV. The overall systematic uncertainty for cross sections 
was estimated to be 5.9-8.3 \%, which was determined by the neutron 
emission rate of the calibration source (5 \%), the 
number of the incident LCS photons (3 \%), and the beam size effect.  

\section{\label{results} Results}
\subsection{Data reduction}

The photoneutron cross section measured with a monochromatic 
photon beam is given by 

\begin{equation}
\sigma=\frac{n_n}{N_{\rm t}N_{\gamma} f \epsilon_n(E_n)}
\label{monochro}
\end{equation}

\noindent
where $n_n$ is the number of neutrons detected with the 
$^3$He counters, $N_{\rm t}$ is the number of target nuclei per unit 
area, $N_{\gamma}$ is the number of incident photons, $f$ is the 
correction factor for a thick-target measurement, and $\epsilon(E_n)$
is the neutron detection efficiency. The correction factor 
is given by $f=(1-e^{-\mu d})/(\mu d)$ with the linear 
attenuation coefficient of photons, $\mu$, and the target thickness, $d$.
The attenuation coefficient was taken by interpolation from \cite{Jaeger68} 
for the average energy of the LCS photons.  The energy spread of the LCS 
photon beam in the full-width at half maximum makes negligible contributions 
($\lesssim 0.2$ \%) to the determination of the correction factor $f$, which 
deviates from unity by no more than 6 \% in the present measurements.

Recently, a methodology was developed to determine cross sections 
for reactions induced by a quasi-monochromatic photon beam \cite{Mohr04}. 
When the photon beam has an energy distribution of $n_{\gamma}(E)$, 
$N_{\gamma}\sigma$ in Eq.~(\ref{monochro}) has to be replaced by 
the integral $\int{n_{\gamma}(E)\sigma(E)dE}$. By expanding the cross 
section $\sigma(E)$ in the Taylor series at the average photon 
energy $E_0$, the first term in the Taylor series $\sigma(E_0)$ 
(the cross section at the average energy) was numerically evaluated 
along with the higher-order terms.  The new methodology determines 
$\sigma(E_0)$ in the energy region of astrophysical importance near threshold  
within 6 \% corrections from the monochromatic approximation (Eq. (3)). 
Photoneutron cross sections presented in this paper are based on this methodology. 

\subsection{Photoneutron cross sections\label{sec-cross}}

\begin{figure}
\includegraphics[scale=0.5]{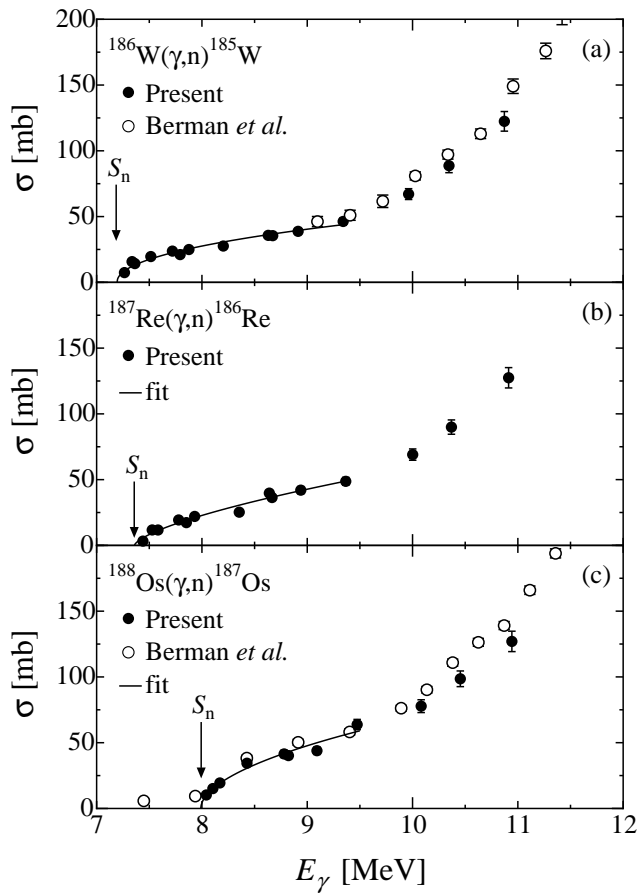}
\caption{Photoneutron cross sections of the reactions
$^{186}$W($\gamma,n$)$^{185}$W (a),
$^{187}$Re($\gamma,n$)$^{186}$Re (b),
and $^{188}$Os($\gamma,n$)$^{187}$Os (c) 
extracted from the present data are plotted with filled circles. 
The previous data for $^{186}$W \cite{Ber69} and $^{188}$Os 
\cite{Ber79} taken with quasi-monochromatic photon beams 
are also shown with open circles for comparison. 
Given with solid line is the best fits to the experimental data 
in Eq. \ref{cr_thres}. 
}
\label{crossfig}
\end{figure}

Photoneutron cross sections measured for $^{186}$W, $^{187}$Re 
and $^{188}$Os are shown in Fig. \ref{crossfig} and Table \ref{crosstab}. 
The contributions from the reactions with the main contaminants 
($^{189}$Os, $^{190}$Os and $^{192}$Os) of the $^{188}$Os target
were estimated from the previous data from Ref. \cite{Ber79}. 
The error bars in Fig. \ref{crossfig} include 
both the statistical and systematic uncertainties. 
Previously, cross section data were taken for $^{186}$W \cite{Ber69} 
and $^{188}$Os \cite{Ber79} with quasi-monochromatic photons 
from positron annihilation in flight, as shown in Fig. \ref{crossfig} 
for comparison. In addition, data for $^{186}$W and $^{187}$Re 
were taken with bremsstrahlung \cite{Sonn03,Mull04,gory73}. 
In Ref. \cite{gory73}, yield curves obtained in small 
increments of the electron beam energy with 1-MeV spacing were converted 
to cross sections through an unfolding procedure based on the Penfold-Leiss 
method \cite{Penf59}. As mentioned in Ref. \cite{gory73}, it is  
a known fact \cite{Tikh63} that cross sections are not obtained correctly 
because of {\it swings} of the solution, 
though uncertainties resulting from the swings may be reduced 
significantly.

The present $^{188}$Os data are in reasonable agreement with the 
previous data \cite{Ber79} except at the low energy where 
the previous data exhibit non-vanishing cross sections even below the 
neutron threshold of 7.99 MeV. The non-vanishing cross 
sections may be attributed to {\em leftover} in the subtraction of  
contributions of the positron bremsstrahlung admixed with the positron annihilation 
photons. 

The ($\gamma,n$) cross section exhibits the threshold 
behavior \cite{Wign48,Brei58} of

\begin{equation}
\sigma(E)=\sigma_0\left(\frac{E_{\gamma}-S_n}{S_n}\right)^{p}.
\label{cr_thres}
\end{equation}

\noindent
Here, $p$ is related to the neutron orbital angular momentum $\ell$ 
through $p=\ell+1/2$. The values of $p=0.5$ and 1.5 are  
expected for the s- and p-wave neutron decays, respectively. The best 
fits to the present experimental data gave $p=0.47$ and $\sigma_0=78$ 
mb for the $^{186}$W($\gamma,n$) reaction, $p=0.67$ and $\sigma_0=117$ 
mb for the $^{187}$Re($\gamma,n$) reaction, and $p=0.53$ and 
$\sigma_0=143$ mb for the $^{188}$Os($\gamma,n$) reaction. 
The results show a rather pure s-wave character for the $^{186}$W 
and $^{188}$Os($\gamma,n$) reactions, 
and suggest an admixture of p- and/or d-wave neutron emissions following the 
$E1$ excitation of $^{187}$Re. The cross-section parametrization 
(Eq. (\ref{cr_thres})) was also made in the bremsstrahlung measurements 
\cite{Sonn03,Mull04}; the results are in rough agreement with our 
new experimental data.

\begin{table}
\caption{Photoneutron cross sections of the reactions 
$^{186}$W($\gamma,n$)$^{185}$W, $^{187}$Re($\gamma,n$)$^{186}$Re, 
and $^{188}$Os($\gamma,n$)$^{187}$Os. $E_{\rm ph}^{\rm av}$ is the average 
energy of the LCS photon beam. Cross sections are given as 
$\sigma\pm\Delta\sigma({\rm stat.})\pm\Delta\sigma
({\rm syst.})$ in units of mb where $\Delta\sigma({\rm stat.})$ and 
$\Delta\sigma({\rm syst.})$ represent the statistical and systematic 
uncertainties.}
\label{crosstab}
\begin{ruledtabular}
\begin{tabular}{cccccc}
Nucleus& $E_{\rm ph}^{\rm av}$ & 
$\sigma$ & $\Delta\sigma$(stat.)
& $\Delta\sigma$(syst.) $^{\rm a}$\\
&(MeV)&(mb)&(mb)&(mb)\\
\hline
$^{186}$W&7.26&7.40&1.35&0.43\\
&7.33&15.7&1.1&1.0\\
&7.36&14.3&0.5&0.9\\
&7.51&19.6&0.6&1.1\\
&7.72&23.7&0.7&1.4\\
&7.79&21.1&0.8&1.3\\
&7.88&25.0&0.6&1.5\\
&8.20&27.6&0.9&1.6\\
&8.63&35.7&0.8&2.1\\
&8.67&35.5&0.9&2.1\\
&8.91&38.8&0.7&2.3\\
&9.34&46.3&0.8&2.7\\
&9.96&67.1&1.3&3.9\\
&10.35&88.7&1.2&5.2\\
&10.87&122&2&7\\
\\
$^{187}$Re&7.44&3.27&1.46&0.21\\
&7.53&11.7&0.8&0.7\\
&7.58&11.7&0.7&0.7\\
&7.78&19.3&0.8&1.3\\
&7.85&17.3&0.9&1.0\\
&7.93&22.0&0.8&1.3\\
&8.35&25.3&0.8&1.5\\
&8.64&39.8&0.9&2.3\\
&8.67&36.4&1.0&2.1\\
&8.94&42.0&1.1&2.4\\
&9.37&48.7&0.9&2.8\\
&10.00&69.0&1.5&4.0\\
&10.37&90.0&1.4&5.2\\
&10.91&127&2&7\\
\\
$^{188}$Os&8.04&10.2&1.5&0.9\\
&8.10&15.1&1.6&1.0\\
&8.17&19.4&1.1&1.2\\
&8.43&34.4&1.2&2.0\\
&8.78&41.5&1.3&2.5\\
&8.82&40.2&1.2&2.3\\
&9.09&43.9&1.1&2.6\\
&9.47&63.8&1.1&3.7\\
&10.08&77.8&1.8&4.5\\
&10.45&98.6&1.6&5.8\\
&10.94&127&2&7\\
\end{tabular}
\end{ruledtabular}
a The uncertainty includes those associated with the neutron detection efficiency (5 \%), the photon flux (3 \%), and the beam size effect (1 - 6 \%) added in quadrature.  
\end{table}

\section{Comparison with theory}

\subsection{Theoretical framework}

The cross sections measured in the present work are now compared with the
predictions of the Hauser-Feshbach (HF) compound nucleus theory \cite{hau52,Holm76}. 
The uncertainties involved in HF cross section
calculation are known not to be related to the theory of compound nucleus emission
itself, but rather to the uncertainties associated with the evaluation of the nuclear
properties entering the calculation of the transmission coefficients. 
It is therefore 
of prime importance to compare the effects of different nuclear inputs to estimate
the reliability and accuracy of the predictions, especially when considering the reverse
rates, i.e, in the present case, the radiative neutron capture rate which might be
sensitive to different input parameters than the rate measured. 
In the present work, the nuclear level densities are derived from two models, either
the widely used back-shifted Fermi gas (BSFG) model based on the global
parametrization of \cite{Gori02} or the microscopic calculations taking into
account the discrete structure of the single-particle spectra associated with
Hartree-Fock+BCS (HFBCS) potentials \cite{dem01}. This model has the advantage of
treating  shell, pairing and deformation effects consistently, and for practical
applications, has been renormalized on existing experimental information (low-lying
levels and s-wave neutron resonance spacings whenever available as in the cases
considered here).  The transmission coefficients for particle emission is calculated
either with the so-called JLMB semi-microscopic potential of \cite{bdg01} derived from
the Br\"uckner--Hartree--Fock approximation based on a Reid's hard core nucleon--nucleon
interaction, or with the global phenomenological mass- and energy-dependent potential of
Woods-Saxon type developed by \cite{koning03}.

The photon transmission function of particular interest in photoemission data is
calculated assuming the dominance of dipole transitions in the photon channel. The
electric- and magnetic-dipole (GDR) transition strength functions are usually described
by a Lorentz-type function where the energies and widths are determined by experimental
data, whenever they exist, or by appropriate parametrizations. However, the calculation
of the radiative capture or photoabsorption at low energies (and particularly in stellar
conditions where the excited states of the target nucleus are thermally populated) is
particularly sensitive to the low-energy tail of the GDR of the compound system. The
shape of the GDR is expressed most frequently by a generalized 
energy-dependent-width Lorentzian
function adjusted on low-energy data \cite{ko90}. To test such models,
we consider here the Hybrid model \cite{go98} which couples the GDR Lorentzian
description at high energies with an analytical approximation to the theory of finite
Fermi systems at energies below the neutron separation energy
\cite{ka83}.

In addition to the Hybrid model \cite{go98}, the Quasi-Particle Random
Phase Approximation (QRPA) model of \cite{kh01} is also considered here for estimating
the photon transmission coefficients. These QRPA calculations are self-consistently built
on a ground state derived with the HFBCS approximation. The final $E1$-strength
functions is obtained by folding the QRPA strength with a Lorentzian function to account
for the damping of collective motions and the deformation effects. This global
calculation based on the SLy4 Skyrme interaction has been shown to reproduce relatively
well photoabsorption and average resonance capture data at low
energies \cite{Utsu03,kh01}.
Note that both the Hybrid and QRPA models differ not only in the predictions of the
position and width of the GDR, but also in the energy dependence of its tail, which is an
important quantity to derive the reaction rate. Since we are here mainly concerned with
the GDR tail at low energies and the prediction of the reverse neutron capture cross
section, both $E1$-strength functions are renormalized on the available experimental
information on the position of the GDR peak and the corresponding maximum absorption
cross section. In the case of the HFBCS+QRPA model, this adjustment is achieved within
the folding procedure introduced to account for damping and deformation effects.

\subsection{Comparison between experimental and theoretical rates}

\begin{figure}
\includegraphics[scale=0.5]{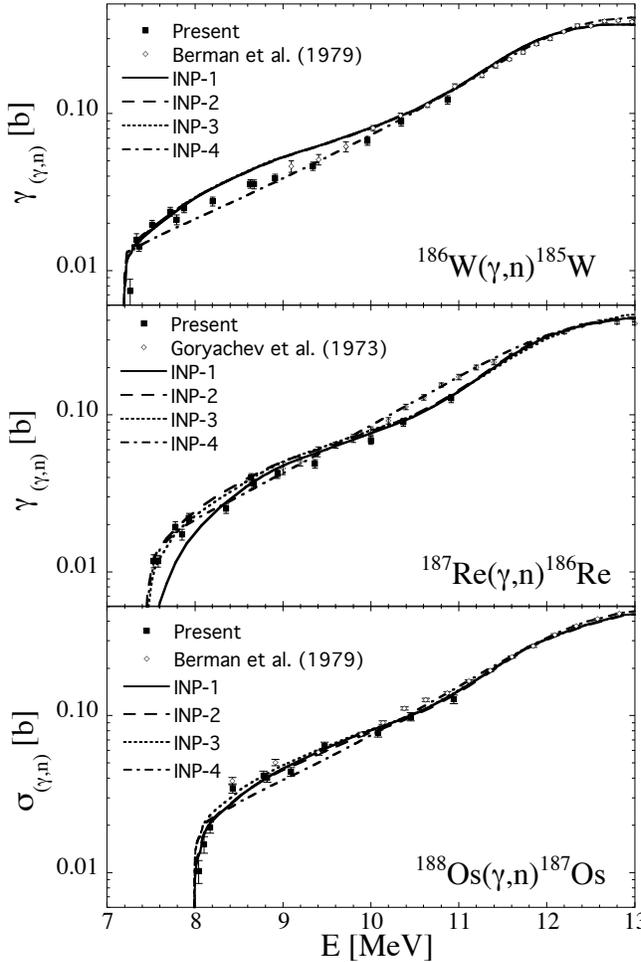}
\caption{Comparison between measured and calculated photoneutron cross sections 
for $^{186}$W($\gamma,n$) (upper panel),
$^{187}$Re($\gamma,n$) (middle panel)
and $^{188}$Os($\gamma,n$) (lower panel). 
The different theoretical predictions correspond to the input defined in Table 
\ref{tab_hf}. 
}
\label{fig_gn}
\end{figure}

The final nuclear inputs considered in the present analysis are summarized in
Table~\ref{tab_hf}. Four different sets are used to estimate the photoemission cross
section as well as the reverse radiative neutron capture cross sections. The comparison
between these 4 sets allow us to estimate the sensitivity of the cross sections to the
various input quantities, but also the uncertainties affecting the
final prediction of the neutron capture rate of astrophysical interest. In
Fig.~\ref{fig_gn}, our new experimental data are compared with the theoretical
photoemission cross sections. Also shown are previous measurements obtained in the
vicinity of the GDR peak energy \cite{Ber79,gory73}. Most of the calculations agree
relatively well with experimental data, down to  energies close to the neutron threshold.
The neutron optical potential and nuclear level densities influence the photoneutron
cross section only in a small energy range of no more than 1 MeV
above the neutron threshold, so that the global behavior of cross section is almost
entirely dictated by the $E1$-strength. Some specific comments can be made for each
reaction:
\begin{itemize}
\item In the 
$^{186}$W($\gamma,n$)$^{185}$W case, the HFBCS+QRPA model predicts some extra strength
at energies around 7.5-10 MeV with respect to the Hybrid model. This extra strength is
clearly seen experimentally below 8 MeV but not above. However, all models overestimate
the 7mb cross section at $E=7.26$~MeV. 

\item In the case of the $^{187}$Re($\gamma,n$)$^{186}$Re reaction, the low-energy data
can only be reproduced when adopting the BSFG model of nuclear level densities. The
microscopic HFBCS-based model fails to describe the fast rise of the cross section at the
neutron threshold. Around 11 MeV, the Hybrid and HFBCS+QRPA strength predict a relatively
different cross section, the former one being compatible with the Goryachev et al.
\cite{gory73} and the later with our more accurate measurements. 

\item As far as $^{188}$Os($\gamma,n$)$^{187}$Os is concerned, all HF calculations
reproduce relatively well the data, though the Hybrid model gives a lower
cross section in the 8.5-10 MeV energy range. 
\end{itemize}

\begin{table*}
\caption{\label{tab_hf}
Overview of the 4 sets of nuclear ingredients adopted in
the Hauser-Feshbach calculations.}
\begin{ruledtabular}
\begin{tabular}{llll}
{Label}& {Level
density}&{$\gamma$-strength} &{Optical potential}\\
\hline
INP-1 &
HFBCS \cite{dem01} & HFBCS+QRPA \cite{kh01} & JLMB \cite{bdg01}\\
INP-2 &
BSFG \protect\cite{Gori02} & HFBCS+QRPA \cite{kh01} & JLMB \cite{bdg01}\\
INP-3 &
BSFG \protect\cite{Gori02} & HFBCS+QRPA \cite{kh01} &  Woods-Saxon
\cite{koning03}\\ 
INP-4 &
BSFG \protect\cite{Gori02}  & Hybrid model\cite{go98} & JLMB \cite{bdg01}\\
\end{tabular}
\end{ruledtabular}
\end{table*}

\subsection{Determination of the neutron capture cross sections}

We now estimate the stellar Maxwellian-averaged neutron capture cross section $\langle
\sigma\rangle^*$ of astrophysics interest on the basis of the calculations presented
above, i.e constrained by the reverse photo-disintegration rate compatible with the new
measurements. 
It should be recalled that the neutron capture cross section at energies of a few tens of
keV is mainly sensitive to photon transmission coefficient at an energy close to and 
even below the
neutron separation energy. For this reason, only the calculations
that reproduce the photoemission rate at the neutron separation energy relatively well
are retained. This leads us to reject the calculation `INP-1' for the
$^{187}$Re($\gamma,n$)$^{186}$Re reaction. 

\begin{figure}
\includegraphics[scale=0.35]{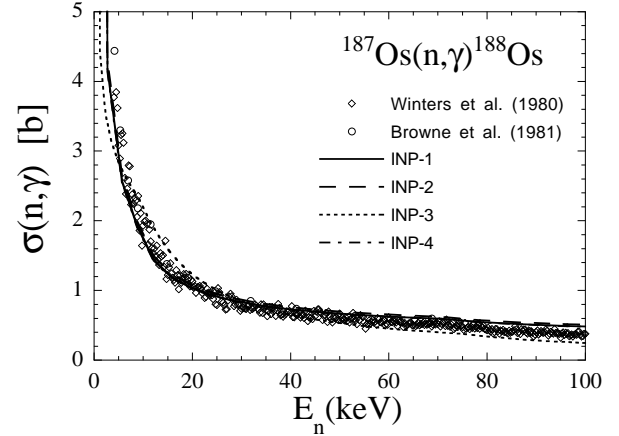}
\caption{Comparison between calculated and measured (open diamonds from \cite{Wint80} 
and open circles from \cite{Brow81}) neutron capture cross sections on $^{187}$Os. 
}
\label{fig_187osng}
\end{figure}

In the particular case of the stable $^{187}$Os target, direct experimental data are
available for the $^{187}$Os($n,\gamma$)$^{188}$Os reaction \cite{Wint80,Brow81} and 
used as additional constraints on the nuclear ingredients, namely the combination of the
nuclear level density and optical potential. The resulting cross sections obtained with
the 4 sets of nuclear inputs given in Table~\ref{tab_hf} are shown in
Fig.~\ref{fig_187osng} and seen to agree relatively well with experimental data, except
in the specific `INP-3' calculation where the use of the global optical potential
\cite{koning03} give rise to a cross section with an energy dependence relatively
different from the one measured.

\begin{figure}
\includegraphics[scale=0.5]{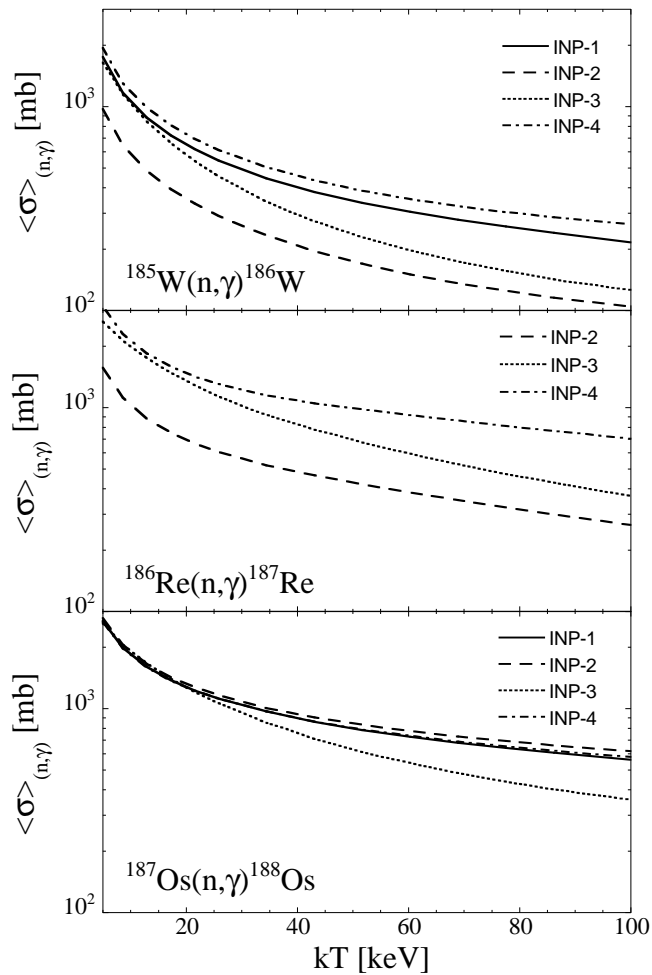}
\caption{Stellar Maxwellian-averaged neutron capture cross sections for $^{185}$W (upper panel), 
$^{186}$Re (middle panel), and $^{187}$Os (lower panel) calculated on the basis of the 
HF input defined in Table \ref{tab_hf}. 
}
\label{fig_ng}
\end{figure}

The stellar Maxwellian-averaged neutron capture cross sections for the three
reactions studied here are shown in Fig.~\ref{fig_ng}. It is evident that
although the photoneutron cross section is relatively insensitive to the nuclear level
density and neutron-nucleus optical potential, particularly in the
$^{186}$W($\gamma,n$)$^{185}$W case, these quantities can affect the reverse rate
significantly. If we characterize the remaining uncertainty affecting the prediction of
the neutron capture rate by the ratio between the upper and lower limits obtained in
Fig.~\ref{fig_ng}, we find at an energy of 25~keV a factor of 1.9 for
the $^{185}$W($n,\gamma$)$^{186}$W reaction, of 2.1 for 
$^{186}$Re($n,\gamma$)$^{187}$Re and only 1.1 for 
$^{187}$Os($n,\gamma$)$^{188}$Os. In the last case, the small error
bars arise from the additional constraints made available through the
experimental $^{187}$Os($n,\gamma$)$^{188}$Os cross section.

\section{Implications to the Re-Os chronometry}

At stellar temperatures relevant to the s-process nucleosynthesis 
($T\simeq$1-3$\times10^8$~K), the $^{187}$Os first excited state at
9.75keV is strongly populated and can significantly
affect the estimate of the stellar neutron capture rate on $^{187}$Os. 
The correction to the cross sections due to the neutron capture on the 9.75 keV 
state is introduced by $F_{\sigma}$ in Eq. (\ref{eq_fsig}).

\begin{figure}
\includegraphics[scale=0.35]{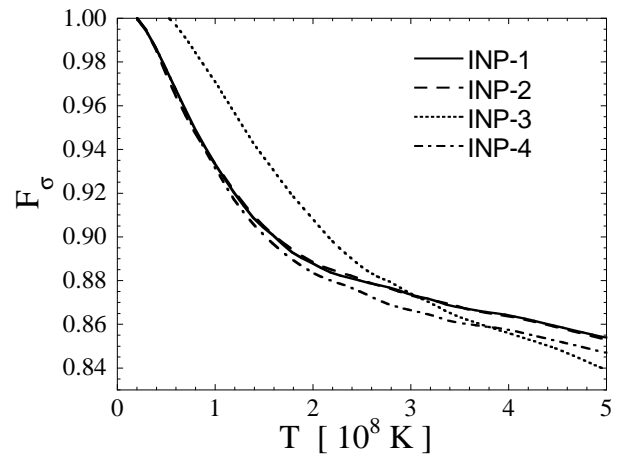}
\caption{Prediction of the $F_{\sigma}$ factor on the basis of the different HF 
calculations defined in Table \ref{tab_hf} and compared in Figs. \ref{fig_gn} and 
\ref{fig_187osng} with experimental data. 
}
\label{fig_fsigma}
\end{figure}

On the basis of the present photoneutron data and the calculations, 
the temperature dependence of the $F_\sigma$ factor
has been re-estimated (Fig.~\ref{fig_fsigma} and Table~\ref{tab_fsig}). At the 
s-process temperature of  $3\times10^8$ K, all calculations converge to the $F_\sigma$ 
value of about 0.87. However, at lower temperatures, the model INP-3 
based on the global
phenomenological optical potential predicts significantly larger 
$F_\sigma$ values. This difference mainly originates from deviations seen in the laboratory
cross section (cf. Fig.~\ref{fig_187osng}) and should therefore be given a lower
credibility. The present calculation agrees relatively well with the value of
$0.80\leq F_\sigma\leq 0.83$ at $kT$=30 keV from Ref. \cite{Hers83}, 
but not with the value of $1\leq F_\sigma\leq1.15$ from Ref. \cite{Wint82}.

\begin{table}
\caption{Comparison between the $F_{\sigma}$ values obtained in the present work
(including INP-3 calculation) and those of Ref. \cite{Holm76} for different temperatures
$kT$ (expressed in keV).}
\label{tab_fsig}
\begin{ruledtabular}
\begin{tabular}{ccc}
$kT$&  Present & Ref.\cite{Holm76} \\
\hline
12 & 0.901 -- 0.937 & 0.867 \\
20 & 0.879 -- 0.886 & 0.839 \\
25 & 0.866 -- 0.874 & 0.830 \\
30 & 0.859 -- 0.867 & 0.820 \\
52 & 0.822 -- 0.847 & 0.813 \\
\end{tabular}
\end{ruledtabular}
\end{table}

We now consider implications of the $F_{\sigma}$ values constrained by the present 
study to the Re-Os cosmochronology. The most tantalizing aspect of the Re-Os 
chronology is that it requires a rather detailed model of the chemical evolution of 
the Galaxy. 
Since constructing a reasonable model of the Galactic chemical evolution 
is beyond the scope of the present study, we rather focus on the uncertainty 
in the estimate of the age of the Galaxy within a schematic model.
   
We recall here that abundances of elements in the relevant mass region can be 
symbolically expressed as follows:\\

$^{\it 186}Os^{\odot}={^{\it 186}Os^{s}}+{^{\it 186}Os^{p}}$

$^{\it 187}Os^{\odot}={^{\it 187}Os^{s}}+{^{\it 187}Os^{c}}$

$^{\it 187}Re^{\odot}={^{\it 187}Re^{r}}+{^{\it 187}Re^{s}}-{^{\it 187}Re^{c}}$

$^{\it 187}Os^{c}={^{\it 187}Re^{c}}$\\

\noindent
Here, $\odot$, $s$, $p$, $r$, and $c$ represent the solar, s-process, p-process, 
r-process, and cosmoradiogenic origins, respectively. We introduce the following 
approximations:\\

$^{\it 186}Os^{p}=p\times{^{\it 186}}Os^{\odot}$

$^{\it 187}Re^{s}=0$

$^{\it 187}Os^{s}=[F_{\sigma}\langle\sigma\rangle(^{186}$Os)/$\langle\sigma\rangle(^{187}$Os)] 
$^{\it 186}Os^{s}$ (local approximation: Eq. (\ref{localapprox}))\\

\noindent
Following the recent p-process calculations \cite{Arno03}, we 
estimated $p$ as lying within the 0.01-0.04 range, being in 
agreement with $p$=0.02 \cite{Kaep91}. As noted earlier, we 
ignore the possible s-process contribution to $^{187}$Re 
through the branchings at $^{185}$W and $^{186}$Re. 

Thus, the cosmoradiogenic component of $^{187}$Os can be obtained as 

\begin{equation}
^{\it 187}Os^{c}={^{\it 187}Os^{\odot}}-(1-p)
[F_{\sigma}\langle\sigma\rangle(^{186}{\rm Os})/\langle\sigma\rangle(^{187}{\rm Os})] 
{^{\it 186}Os^{\odot}}.
\label{eq:187osc}
\end{equation}

A consideration of the simplest model of a closed system would lead us 
to the assumption that the evolution of $^{187}$Re can be effectively described by

\begin{equation}
\frac{d{^{\it 187}Re(t)}}{dt}=-\lambda{^{eff}_\beta}{^{\it 187}Re}(t)+Y(t)
\label{eq:diff}
\end{equation}

\noindent
where $\lambda_\beta^{eff}$ is the effective $\beta$-decay rate of $^{187}$Re 
in consideration of some enhancement by {\it astration} from the laboratory 
decay rate of $\lambda_{\beta}$ = ln2/(41.6 Gyr), whereas $Y(t)$ term represents 
the net r-process yield. As in Ref. \cite{Kaep91}, we adopt a simple form of 
$Y(t)=y\ {\rm exp}(-\lambda t)$ where $\lambda$ is a free parameter. Thus, we have

\begin{equation}
{^{\it 187}Re}(t) = \frac{y [e^{- \lambda_{\beta}^{eff}t}-
e^{- \lambda t}]}{\lambda - \lambda_{\beta}^{eff}}.
\label{eq:solut}
\end{equation}

\noindent
Using $d^{\it 187}Os^c(t)/dt$ = + $\lambda_\beta^{eff}{^{\it 187}}Re(t)$, 
the abundance ratio between $^{\it 187}Os^{c}(t)$ and $^{\it 187}Re(t)$ can be 
obtained as 

\begin{equation}
\frac{{^{\it 187}Os}^{c}(t)}{{^{\it 187}Re}(t)} = \frac{B}{A}
\label{eq:ratio}
\end{equation}

\noindent
where

\begin{equation}
A=e^{-\lambda_{\beta}^{eff} t}-e^{-\lambda t}
\label{eq:A}
\end{equation}

 \begin{equation}
B=[1-e^{-\lambda_{\beta}^{eff} t}]-[1-e^{-\lambda t}]\lambda_\beta^{eff}/
\lambda.
\label{eq:B}
\end{equation}

\noindent
The abundance ratio in Eq.~(\ref{eq:ratio}) at 4.55 Gyr ago is matched with that from Eq.~(\ref{eq:187osc}).

For the meteoritic abundance of cosmoradiogenic $^{\it 187}Os^{c}$ relative 
to $^{\it 187}Re^{\odot}$, we used the following solar abundances \cite{Faes98}:\\

$^{\it 187}Re^{\odot}$/$^{\it 186}Os^{\odot}$ = 3.51 $\pm$ 0.09

$^{\it 187}Os^{\odot}$/$^{\it 186}Os^{\odot}$ = 0.793 $\pm$ 0.001.\\

\noindent
The ($n,\gamma$) cross section ratios at s-process temperatures, 
$\sigma(^{186}{\rm Os})/\sigma(^{187}{\rm Os})$, were taken from Ref. \cite{Kaep91}. 
Thus, the meteoritic quantity is determined with the present $F_{\sigma}$ 
value being a unique parameter.  As for $\lambda^{eff}_{\beta}$/$\lambda_{\beta}$
is used to calculate the abundance ratio in Eq. (\ref{eq:ratio}) at $t$ = $T_{G}$ 
(the age of the Galaxy) $-4.55$ Gyr, Clayton \cite{Clay88} derived 1.4 from the work 
of Yokoi {\it et al.} \cite{Yoko83}, whereas more recent analyses suggest considerably 
lower values \cite{Taka04}.  We adopt here 1.2 as our standard value for the net 
enhancement by {\it astration} of the $^{187}$Re $\beta$-decay rate.  

Matching conditions in ${^{\it 187}Os}^c$/$^{\it 187}Re^{\odot}$ were 
investigated in the $T_{G}$ range of 11 - 15 Gyr, showing good agreement 
with $0 \lesssim \lambda \lesssim 2$ Gyr$^{-1}$.
We summarize some conclusions obtained under the simplest assumption of 
r-process nucleosynthesis yields varying exponentially in time.

\begin{itemize}
\item 
$F_{\sigma}$ values at typical s-process temperatures are in the range of 
0.86-0.94 (Table \ref{tab_fsig}).  

\item 
The probable range of the differential coefficient $dT_{G}/dF_{\sigma}$ 
is $-$(5.0-12.8) Gyr.

\item
Consequently, the remaining uncertainty of $T_{G}$ that stems from that of 
$F_{\sigma}$ values is less than 1 Gyr. When the temperature dependence of 
$\langle\sigma\rangle(^{186}{\rm Os})/\langle\sigma\rangle(^{187}{\rm Os})$ is considered along with 
that of $F_{\sigma}$, the uncertainty in $T_{G}$ is approximately halved.  
\end{itemize}

We note here that the model of Yokoi {\it et al.} \cite{Yoko83} cannot be 
reconciled with the present $F_{\sigma}$ data. The model of chemical evolution 
developed there favored $F_{\sigma}$ values much higher than unity. If we 
mimic the results in terms of Eq.~(\ref{eq:diff}), the corresponding values of 
$\lambda$ become negative. This also explains the much larger 
$|dT_{G}/dF_{\sigma}|$
values of up to 100 Gyr (as seen in the slope of Fig.~\ref{eq:A} 
in Ref. \cite{Yoko83}). Finally, it is noted that if we would use the 
age of the universe as derived by the WMAP ($T_{U}$ = 13.7 $\pm$ 0.2 Gyr) 
\cite{WMAP}, and assume $T_{U}$ - $T_{G}$ $\approx$ 1 Gyr, $\lambda$ values 
that are consistent with the present $F_{\sigma}$ value are narrowed in 
the ranges of 0.16-0.46 Gyr$^{-1}$, inclusively, and 0.22-0.34 Gyr$^{-1}$, 
exclusively. The exclusive values are commonly allowed in calculations 
with all possible combinations of $kT$, $F_{\sigma}$, and $T_{G}$.  

\section{Conclusion}
Photoneutron cross sections were measured with accuracy 
for $^{186}$W, $^{187}$Re and $^{188}$Os using quasi-monochromatic photon beams
from laser Compton scattering (LCS) at energies near the neutron thresholds. 
The cross sections were used to
constrain the model parameters in the framework of the Hauser-Feshbach
model. Four different sets of nuclear ingredients were adopted to estimate the
photoneutron cross sections and the reverse radiative neutron capture cross sections.
When no experimental data on the direct ($n,\gamma$) cross section is available, an
accuracy of about a factor of 2 was achieved in the predictions. 
The influence of the neutron capture by the 9.75 keV first excited state in
$^{187}$Os which is substantially populated in stellar plasmas at typical s-process
temperatures has been estimated in connection with the $^{187}$Re-$^{187}$Os
cosmochronology and shown to lead to an increase of the neutron capture rate by a 
factor of about 1.15 at a temperature of $3\times10^8$ K. Uncertainties by about 10\% 
associated with the neutron-nucleus optical potential still affect the stellar 
rate at temperatures between 1 and $2\times10^8$ K.

The correction factor $F_{\sigma}$ to be used in the local approximation 
(Eqs. (\ref{localapprox}) and (\ref{eq_fsig})) was constrained well in 
the present study (Table \ref{tab_fsig}). Based on the simplest assumption of r-process 
nucleosynthesis yields varying exponentially in time, the cosmochronological 
uncertainty in the age of the Galaxy arising from the $F_{\sigma}$ values is 
estimated to be less than 1 Gyr; when the temperature dependences of both 
$\langle\sigma\rangle(^{186}$Os)/$\langle\sigma\rangle(^{187}$Os) and $F_{\sigma}$ are considered, 
the uncertainty is less than 0.5 Gyr.  

\begin{acknowledgments}
H.U. is grateful to Kohji Takahashi for helpful suggestions. 
This work was done within the Konan-ULB joint project and supported by 
the Japan Atomic Energy Research Institute (the REIMEI Research Resource), 
the Japan Private School Promotion Foundation, and the Japan Society for the
Promotion of Science. S.G. is FNRS Research Associate.  

\end{acknowledgments}


\begin{references}
\bibitem{Clay64}
D.D. Clayton, 
Astrophys. J. {\bf 139}, 637 (1964).

\bibitem{AnGr89}
E. Anders and N. Grevesse, 
Geochin. Cosmochim. Acta. {\bf 53}, 197 (1989).  

\bibitem{Yoko83}
K. Yokoi, K. Takahashi, and M. Arnould, 
Astron. Astrophys. {\bf 117}, 65 (1983).

\bibitem{Arno84}
M. Arnould, K. Takahashi, and K. Yokoi, 
Astron. Astrophys. {\bf 137}, 51 (1984).

\bibitem{Taka03}
K. Takahashi, 
Nucl. Phys. {\bf A718}, 325c (2003).

\bibitem{Woos79}
S.E. Woosley and W.A. Fowler, 
Astrophys. J. {\bf 233}, 411 (1979).

\bibitem{Holm76}
J.A. Holmes, S.E. Woosley, W.A. Fowler, and B.A. Zimmerman, 
Atomic Data and Nuclear Data Tables {\bf 18}, 306 (1976).

\bibitem{Fowl73} 
W.A. Fowler, {\it in Explosive Nucleosynthesis} 
ed. N.D. Schramm and W.D. Arnett 
(Austin: University of Texas Press 1973), p.300.

\bibitem{Wint80} 
R.R. Winters, R. L. Macklin, and J. Halperin, 
Phys. Rev. C {\bf 21}, 563 (1980).

\bibitem{Wint82}
R.R. Winters and R.L. Macklin, 
Phys. Rev. C {\bf 25}, 208 (1982).

\bibitem{Hers83}
R.L. Hershberger, R.L. Macklin, M. Balakrishnan, N.W. Hill, and M. T. McEllistrem, 
Phys. Rev. C {\bf 28}, 2249 (1983).

\bibitem{Mack83}
R.L. Macklin, R.R. Winters, N.W. Hill, and J.A. Harvey, 
Astrophys. J. {\bf 274}, 408 (1983).

\bibitem{BrLS76} 
J.C. Browne, G.P. Lamaze, and I.G. Schroder, 
Phys. Rev. C {\bf 14}, 1287 (1976).

\bibitem{Brow81}
J.C. Browne and B.L. Berman, 
Phys. Rev. C {\bf 23}, 1434 (1981).

\bibitem{Wint74}
R.R. Winters, F. K\"appeler, K. Wisshak, B.L. Berman, and J.C. Browne, 
Bull. Am. Phys. Soc. {\bf 24}, 854 (1974).

\bibitem{WiMH87}
R.R. Winters, R.L. Maclin, and R.L. Hershberger, 
Astron. Astrophys. {\bf 171}, 9 (1987).

\bibitem{McEl89}
M.T. McEllistrem, R.R. Winters, R.L. Hershberger, Z. Cao, R.L. Macklin, and N.W. Hill, 
Phys. Rev. C {\bf 40}, 591 (1989).

\bibitem{Ber69}
B.L. Berman, M.A. Kelly, R.L. Bramblett, J.T. Caldwell, H.S. Davis, and S.C. Fultz, 
Phys. Rev. {\bf 185}, 1576 (1969).

\bibitem{Kaep91}
F. K\"{a}ppeler, S. Jaag, Z. Y. Bao, and G. Reffo, 
Astron. Astrophys. {\bf 366}, 605 (1991).

\bibitem{Mohr04}
P. Mohr, T. Shizuma, H. Ueda, S. Goko, A. Makinaga, K. Y. Hara,
T. Hayakawa, Y.-W. Lui, H. Ohgaki, and H. Utsunomiya,
Phys. Rev. C {\bf 69}, 032801(R) (2004).

\bibitem{Ohg91}
H. Ohgaki, S. Sugiyama, T. Yamazaki, T. Mikado, M. Chiwaki, 
K. Yamada, R. Suzuki, T. Noguchi, and T. Tomimasu, 
IEEE Trans. Nucl. Sci. {\bf 38}, 386 (1991).

\bibitem{Utsu03}
H. Utsunomiya, H. Akimune, S. Goko, M. Ohta, H. Ueda, T. Yamagata, 
K. Yamasaki, H. Ohagaki, H. Toyokawa, Y.-W. Lui, T. Hayakawa,
T. Shizuma, E. Khan, and S. Goriely, 
Phys. Rev. C {\bf 67}, 015807 (2003).

\bibitem{Nel85}
W.R. Nelson, H. Hirayama, and W.O. Roger, 
The EGS4 Code Systems, SLAC-Report-265, (1985).

\bibitem{Toyo00}
H. Toyokawa, T. Kii, H. Ohgaki, T. Shima, T. Baba, and Y. Nagai,
IEEE Trans. Nucl. Sci. {\bf 47}, 1954 (2000).

\bibitem{Hara03}
K.Y. Hara, H. Utsunomiya, S. Goko, H. Akimune, T. Yamagata, M. Ohta, 
H. Toyokawa, K. Kudo, A. Uritani, Y. Shibata, Y.-W. Lui, and H. Ohgaki,
Phys. Rev. D {\bf 68}, 072001 (2003).

\bibitem{Jaeger68}
{\it Engineering Compendium on Radiation Shielding}, Vol. 1, 
{\it Shielding Fundamentals and Methods}, edited by R.G. Jaeger, 
E.P. Blizard, A.B. Chilton, M. Grotenhuis, A. H\"onig, 
Th.A. Jaeger, and H.H. Eisenlohr (Springer-Verlag, New York, 1968), p.175. 

\bibitem{Ber79}
B.L. Berman, D.D. Faul, R.A. Alvarez, P. Meyer, and D.L. Olson,
Phys. Rev. C {\bf 19}, 1205 (1979).

\bibitem{Sonn03}
K. Sonnabend, P. Mohr, K. Vogt, A. Zilges, A. Mengoni, 
T. Rauscher, H. Beer, F. K\"{a}ppeler, and R. Gallino,
Astrophys. J. {\bf 583}, 506 (2003).

\bibitem{Mull04}
S. M\"uller, 
diploma thesis, Technische Universit\"at Darmstadt, 2004 (unpublished).

\bibitem{gory73}
A.M. Goryachev, G.N.Zalesnyi, S.F. Semenko, and B.A. Tulupov,
Yad. Fiz. {\bf 17}, 463 (1973).

\bibitem{Penf59}
A.S. Penfold and J.E. Leiss, 
Phys. Rev. {\bf 114}, 1332 (1959).

\bibitem{Tikh63}
A.N. Tikhonov, 
Dokl. Akad. Nauk SSSR {\bf 151}, 501 (1963).

\bibitem{Wign48}
E.P. Wigner, 
Phys. Rev. {\bf 73} (1948) 1002.

\bibitem{Brei58} 
G. Breit, 
\PR{107}{1612} (1958).

\bibitem{hau52}
W. Hauser and H. Feshbach, 
Phys. Rev. {\bf 87}, 366 (1952).

\bibitem{Gori02}
S. Goriely, 
J. Nucl. Science and Technology (2002), Supp. 2 (Ed. K. Shibata), 536. 

\bibitem{dem01}
P. Demetriou and S. Goriely, 
\NPA{A695}{95} (2001).

\bibitem{bdg01}
E. Bauge, J.P. Delaroche, and M. Girod, 
\PR{C63}{024607} (2001).

\bibitem{koning03}
A.J. Koning and J.P. Delaroche, 
\NPA{A713}{231} (2003).

\bibitem{ko90}
J. Kopecky and M. Uhl, 
\PR{C41}{1941}  (1990).

\bibitem{go98}
S. Goriely, 
\PL{B436}{10} (1998).

\bibitem{ka83}
S.G. Kadmenskii, V.P. Markushev, and V.I. Furman, 
Sov. J. Nucl. Phys. {\bf 37}, 165 (1983).

\bibitem{kh01}
S. Goriely and E. Khan, \NPA{A706}{217} (2002).

\bibitem{Arno03}
M. Arnould and S. Goriely, 
Phys. Rep. {\bf 384}, 1 (2003).

\bibitem{Faes98}
T. Faestermann, 
Proc. 9th Workshop on Nuclear Astrophysics, 
1998, ed. N. Hillebrandt and E. M\"uller, MPA/P12, p.172.

\bibitem{Clay88}
D.D. Clayton, 
Mon. Not. R. Astr. Soc. {\bf 234}, 1 (1988).

\bibitem{Taka04}
K. Takahashi, 
private communication.

\bibitem{WMAP}
D.N. Spergel, L. Verde, H.V. Peiris, E. Komatsu, M.R. Nolta, C.L. Bennett, 
M. Halpern, G. Hinshaw, N. Jarosik, A. Kogut, M. Limon, S.S. Meyer, 
L. Page, G.S. Tucker, J.L. Weiland, E. Wollack, and E.L. Wright, 
Astrophys. J. Supp. {\bf 148}, 175 (2003).

\end{references}
\end{document}